# Single to quadruple quantum dots with tunable tunnel couplings

T. Takakura[1], A. Noiri[1], T. Obata[1], T. Otsuka[1,2], J. Yoneda[1], K. Yoshida[1], and S. Tarucha[1,2]

[1]*Department of Applied Physics, University of Tokyo, 7-3-1 Hongo, Bunkyo-ku, Tokyo 113-8656, Japan*
[2]*RIKEN, Center for Emergent Matter Science, 3-1 Wako-shi, Saitama 351-0198, Japan*

We prepare a gate-defined quadruple quantum dot to study the gate-tunability of single to quadruple quantum dots with finite inter-dot tunnel couplings. The measured charging energies of various double dots suggest that the dot size is governed by gate geometry. For the triple and quadruple dots we study gate-tunable inter-dot tunnel couplings. Particularly for the triple dot we find that the effective tunnel coupling between side dots significantly depends on the alignment of the center dot potential. These results imply that the present quadruple dot device has gate performance relevant for implementing spin-based four-qubit systems with controllable exchange couplings.

Quantum dots (QDs) are artificial structures fabricated in semiconductors in which electrons are confined within the size of their de-Broglie wave length, typically tens of nanometers. Since QDs can trap single electrons isolated from the environment and these electron states are precisely controlled, they are attractive systems for both basic research of electron interaction and applications to quantum information processing. Recently several challenging experiments have demonstrated coherent manipulation of electron spins[1-4] following the proposal of electron-spin-based quantum computation [5].

We previously demonstrated two spin-1/2 qubits and exchange control with a double quantum dot (DQD) with a micro-magnet (MM) [4] and proposed a triple QD (TQD) with a MM suitable for implementing three spin-1/2 qubits [6]. Extending the number of qubits is an important step toward realization of quantum computation. Several types of few-electron TQDs have been demonstrated in recent years [7-9]. As for quadruple QDs (QQDs), some systems consisting simply of two capacitively coupled DQDs have been studied [10-13]. In these devices each DQD is used as a charge qubit [10, 11] or a singlet-triplet spin qubit [12, 13] and the capacitive coupling between the two DQDs has been used to perform conditional operations between the qubits. However, no QQDs having finite tunnel couplings between all the neighboring dots have ever been fabricated. Furthermore, integration with a MM favors multiple QDs in a linear array. In this Letter we fabricated collinear QQDs with inter-dot tunnel coupling, which are designed to be fitted with a MM, and demonstrated gate-tunable formation of single, double, triple and quadruple QDs by adjusting gate voltages and observed the effect of inter-dot tunneling in the stability diagram.

Figure 1(a) shows a scanning electron micrograph of our device. The geometry of the surface gate electrodes are designed by using the numerical simulation[14] of electrostatic potential to create four dots in a row. The gate-defined QQD is formed in a 100-nm deep two-dimensional electron gas (2DEG) at a GaAs/AlGaAs hetero-interface with a capping gate on top to effectively reduce the 2DEG density. All experiments to identify the charge states of the fabricated devices were performed at a bath temperature of 50 mK.

Initially we applied appropriate gate voltages to form three different DQDs A-B, A-BC, and A-BCD as shown pictorially in Fig. 1 (b), (c), and (d), respectively and measured the conductance to quantitatively characterize the DQDs. Here each QD is labeled QD A to D, as shown in Fig. 1(a). Although QD B and C are adjacent to the reservoir, the channels between the gates T and TL, and T and TR are set to be pinched off. The charge stability diagram of each DQD measured as a function of two gate voltages is shown in Fig. 1(b) to (d). The source-drain bias voltage $V_{SD}$ = 100 μV. Formation of the DQD is distinguished by observation of two sets of Coulomb peaks with different slopes in the stability diagram. The difference of the slopes results from the difference in the capacitive couplings of the dot to the respective gate electrode. The Coulomb peaks are more visible in (b) and (c) than in (d) because the inter-dot tunnel coupling and the tunnel coupling of the DQD to the reservoir are stronger in (b) and (c). In (d), on the other hand, each cross-point of two Coulomb peaks is observed as two separate triple points, indicating the tunnel couplings are all weak.

To quantitatively characterize the sizes of the DQDs, we estimated the charging energy of each dot from the high-bias stability diagram (not shown). When $V_{SD}$ across the DQD is increased, the triple points evolve into bias triangles whose size increases in proportion with the magnitude of $V_{SD}$ [15]. Using the width of the triangle along the diagonal line as a measure of energy, we estimated the charging energy of the two QDs, QD α and QD β, in each DQD, $E_{C\alpha(\beta)}$, from the interval between Coulomb peaks. The obtained charging energy ranges from 1 to 3 meV.

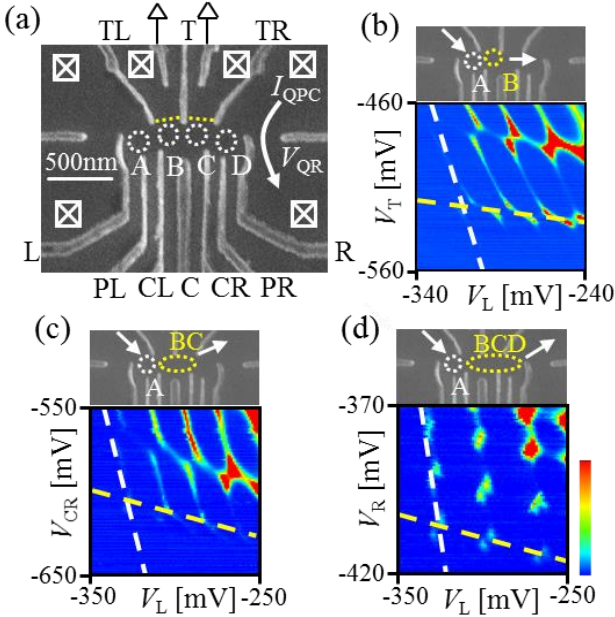

FIG. 1. Device structure and several DQD configurations. (a) Scanning electron micrograph of the QQD device. White circles indicate the location of the dots. Ohmic contacts are shown as white boxes. A white arrow outlines the current flowing through the right QPC charge sensor. The QPCs shown as yellow dotted lines are pinched off and the two gate electrodes on both sides of the T gate are grounded throughout the present experiment. (b), (c), and (d) Charge stability diagrams of DQD A-B, A-BC, and A-BCD measured in electron transport as a function of two gate voltages, respectively. White (yellow) dashed lines emphasize the charge transition in the left (right) dot. Upper panels illustrate the position of each dot labeled QD A to D. Color bar shows 0 pA (blue) to 100, 50, and 20 pA (red) in (b), (c), and (d), respectively.

If the inter-dot capacitance $C_m$ is much smaller than the total capacitance of each dot, the charging energy is represented just like that of a single QD (SQD) as $E_{C\alpha} = e^2 C_\beta / (C_\alpha C_\beta - C_m^2) \cong e^2 / C_\alpha$ with the capacitance of QD $\alpha$ ($\beta$), $C_{\alpha(\beta)}$, where $\alpha$ ($\beta$) denotes the respective set of QD A to D. Table I shows the estimated capacitances of DQD $\alpha$-$\beta$. For example, comparing the DQD A-B with DQD A-BCD, the capacitance of the right dot or QD BCD of DQD A-BCD is four times larger than that of QD B of DQD A-B while the capacitance of the left dot or QD A is only 1.5 times larger. Since the capacitance of QD is more or less proportional to its size, this result indicates that we could properly change the size of the right dot of the DQD. The smaller capacitance of QD A in the DQD A-B than in the DQD A-BCD is probably due to the larger negative voltage applied to the T gate. Comparing the capacitances in all DQDs, we derive the ratio of the capacitances as $C_A \simeq 2C_B \simeq 2C_C \simeq C_D$. Therefore we conclude that the DQDs are formed as intended with controlled size.

| Formation | Left dot (aF) | Right dot (aF) |
|---|---|---|
| A-B | 100 | 50 |
| A-BC | 110 | 110 |
| A-BCD | 150 | 200 |
| AB-CD | 130 | 150 |

TABLE I. Capacitance of each dot in various DQD formation

We then switched the measurement from conductance to charge sensing with quantum point contacts (QPCs) located on each side of the QQD. Charge sensing is a powerful tool to derive the charge stability diagram in multiple QDs, where the condition for elastic transport is severely restricted. We measured the right QPC current $I_{QPC}$ as a function of the left side-gate voltage $V_L$. The QPC conductance was kept sensitive enough to detect the change in the charge state. We adjusted the gate voltages to tune the inter-dot tunnel couplings in various ways such that the dot configuration is changed from single to quadruple QD. In these measurements $V_{SD}$ is set to be very small ($\simeq 0~\mu V$) to detect the ground states.

Figures 2(a), and (b) indicate the charge stability diagrams in the numerical derivative $dI_{QPC}/dV_R$ versus $V_L - V_R$ showing the formation of SQD ABCD, and DQD AB-CD, respectively. Each dark line originates from an abrupt jump of QPC current, which corresponds to single electron charging of one of the QDs. The SQD ABCD is featured by completely parallel charging lines. In this case the inter-dot tunnel couplings are so large that all dots are merged to form a single large dot. In Fig. 2(b), a typical honeycomb structure for a DQD is observed when more negative voltages are applied to the gates T and C. Hence the tunnel coupling between QD B and QD C is very weak, so that the DQD AB-CD is formed.

Figure 2(c) shows the stability diagram of the TQD A-BC-D in which the tunnel couplings between QD A and B and between QD C and D are weak. A dark line with the slope of $dV_R/dV_L = -1$ is assigned to the charging of the center dot or QD BC. Here we mention the inter-dot tunneling between QD A and QD D. Blue and red dashed circles indicate the intersections of the charging lines belonging to QD A and QD D. When the inter-dot tunneling is large, electrons are not fully localized and occupy molecular orbitals spread over two QDs, which are revealed as bending of charging lines near the triple points [16]. Generally the inter-dot tunnel coupling becomes small as the gate voltages are made more negative. However, the blue circle regions show the larger anti-crossing than the red circle even though they are in the range of more negative gate voltages. This implies that indirect tunneling occurs via the energy level of the central dot when it comes closer to those of QD A and D as depicted in the insets of Fig. 2(c). We discuss later this tunneling effect in more detail.

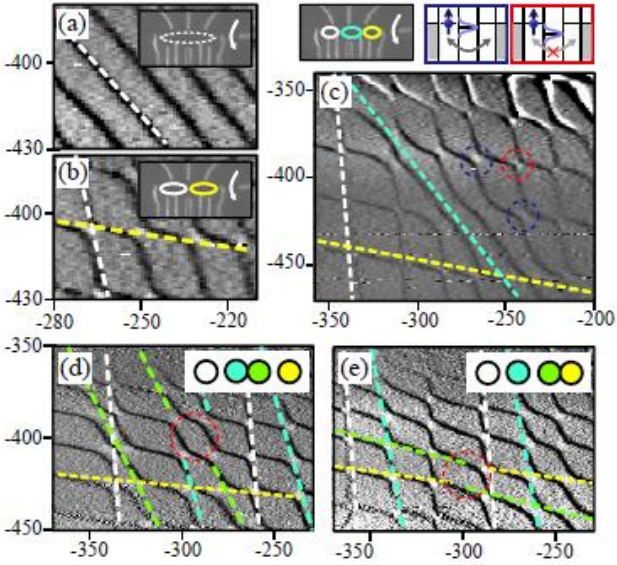

FIG. 2. Charge stability diagrams in the numerical derivative $dI_{QPC}/dV_R$ as a function of $V_L$ and $V_R$ for the formation of single to quadruple QDs: SQD ABCD (a), DQD AB-CD (b), TQD A-BC-D (c), QQD A-B-C-D with two different inter-dot couplings (d) and (e). The dot position is illustrated in the device photo in (a) to (c) and only schematically shown for the QQD with two different gate voltage conditions in (d) and (e). The upper center, and right panels indicate the electrochemical potential ladders corresponding to the situations at the blue, and red circle in the TQD stability diagram in (c). For the QQD, red dashed circles highlight the large anti-crossings between the charging lines of QD B and QD C (d) and QD C and QD D (e), respectively.

We show the charge stability diagrams of QQD configurations in Figs. 2(d) and 2(e). Each dot is separated from its neighbors but there are still a finite inter-dot tunnel coupling in between. Formation of the QQD is confirmed by counting the number of sets of charging lines with different slopes and the size of anti-crossing of two different charging lines, which is larger for the more closely spaced two dots. The inter-dot tunnel couplings are tunable by the gate voltages. Figure 2(d) indicates the case for strong inter-dot coupling between QD B and C. The charging lines of QD B, and C in cyan, and green, respectively are almost inseparable in the less negative gate voltage region, but can be assigned to different dots in the more negative voltage region. On the other hand, Fig. 2(e) is the case for strong coupling between QD C and D. Here large anti-crossings between the charging lines of QD C and D in green and yellow respectively, are observed. From the results of Fig. 2(a) to (e) we confirm that we are able to tune the inter-dot couplings in the QQD configurations as well as form single to quadruple QDs in a single device.

As we mentioned previously, for QD A-BC-D in Fig. 2(c) a finite tunnel coupling between QD A and QD D is observed even though they are spatially separated with the center dot QD BC in between. At this condition the electrochemical potential of the center dot mediates tunneling when it is close to the electrochemical potentials of the side dots [16].

To be more quantitative, we establish a model of the tunnel-coupled TQD as follows (see supplementary material for details). First of all we consider a TQD consisting of three well-separated dots with eigenenergy $E_i$, respectively. We assume $E_1 = E_3 = E$ and $E_2 = E + \varepsilon$ since we pay attention to the anti-crossing between the left and the right dots, where energy levels of the side dots are aligned and that of the center dot is detuned by $\varepsilon$. Then we introduce finite tunnel coupling between neighboring dots, resulting in modified eigenstates and eigenenergies. We define the energy separation due to the indirect tunnel coupling of the two side dots, $E_t = \left(\sqrt{\varepsilon^2 + 8t^2} - |\varepsilon|\right)/2$. The indirect tunnel coupling is revealed as a set of curvatures in the vicinity of the intersection between two charging lines, as shown in Fig. 3(a). We can extract the curvature $t_c$ from the TQD charging diagram and thus obtain $E_t = 2t_c$.

Finally we analyze the data in Fig. 2(c) and estimate the inter-dot coupling between the center and the side dots, which is represented as $t$ in the Hamiltonian of the system. Comparing the width of the bias triangle under finite $V_{SD}$ and the bending of the anti-crossing between the charging lines of the two side dots in the TQD diagram, we evaluate $E_t$ at each anti-crossing. In order to estimate the detuning of the center dot at each anti-crossing, we first calculate the charging energy of the center dot. Judging from Table I, the capacitance of the center dot or QD BC is 100 aF, which can be converted into a charging energy of 1.5 meV. Then comparing the interval of two charging lines of the center dot (the other line is not shown in Fig. 2(c) but exists in the more negative gate voltage region) and the distance between the charging line of the center dot and each anti-crossing point for the charging lines of the two side dots, we derive the detuning value $\varepsilon$ at each point. Figure 3(b) indicates $E_t$ evaluated as a function of $\varepsilon$. The blue curve represents the theoretical fit with $t$ as a parameter. The best fit is obtained for $t = 120\ \mu$eV.

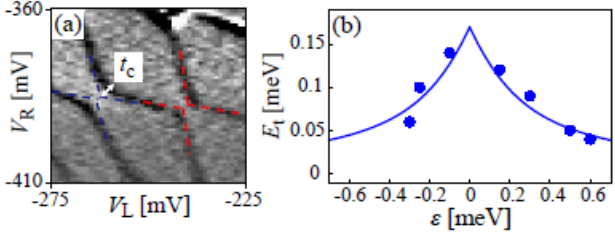

FIG. 3 Indirect tunnel coupling via the center dot in TQD. (a) A zoomed-in diagram taken from Fig. 2(c). Blue (red) dashed lines emphasize the charging lines in the vicinity of strong-coupling (weak-coupling) anti-crossing for the case without inter-dot tunnel coupling. (b) $E_t$ as a function of ε. Each blue point corresponds to the extracted value of $E_t$ at each anti-crossing in Fig. 2(c). The blue curve represents the theoretical fitting.

In conclusion, we fabricated a collinear tunnel-coupled QQD device which is designed to facilitate implementation of a four spin-1/2 qubit system, and studied the gate tunability of single to quadruple QDs with inter-dot gate voltages. We first formed various configurations of DQDs and estimated the charging energy and the capacitance of each dot. The obtained QD capacitance, which is roughly proportional to the dot size, varies as expected from the gate metal geometry. Then in the TQD configuration we qualitatively analyzed the resonant tunneling between both side dots depending on the energy detuning of the center dot from the resonant level. In the QQD configuration we could change the strength of inter-dot tunnel coupling between the two center dots and between the two side dots to study the controllability of tunnel coupling. These results support the fact that the tunnel coupling and therefore the exchange coupling between adjacent dots are tunable with gate voltages. The gate tunability studied here has never been demonstrated for QQDs, indicating that our QQD will be suitable for implementing four spin qubit systems.

We acknowledge Giles Allison for proofreading the English text. Part of this work is financially supported by a Grant-in-Aid for Scientific Research on Innovative Areas (21102003) from the Ministry of Education, Culture, Sports, Science and Technology, Japan, the Funding Program for World-Leading Innovative R&D on Science and Technology (FIRST), and the IARPA project "Multi-Qubit Coherent Operations" through Copenhagen University. TT and JY acknowledge support from JSPS Research Fellowships for Young Scientists. AN acknowledge support from Advanced Leading Graduate Course for Photon Science (ALPS). T. Otsuka acknowledges financial support from the Grant-in-Aid for Research Activity start-up and Young Scientists and T. Obata acknowledges financial support from JSPS Grant-in-Aid for Young Scientists (B) Grant Number 24710148.

Supplementary material:

Single to quadruple quantum dots with tunable tunnel couplings

T. Takakura, A. Noiri, T. Obata, T. Otsuka, J. Yoneda, K. Yoshida, and S. Tarucha

A. Supplementary text

B. Supplementary figures

A. Supplementary text

We establish a theoretical model of the tunnel-coupled TQD in order to analyze the indirect tunnel coupling between the left and the right dots via the energy level of the center dot. First of all we consider a TQD consisting of three well-separated dots. The system is described by a Hamiltonian,

$$H_0|\phi_i\rangle = E_i|\phi_i\rangle \quad (A.1)$$

where $|\phi_i\rangle$ and $E_i$ are the eigenstate and the eigenvalue of the i-th dot (i = 1, 2, 3). We assume $E_1 = E_3 = E$ and $E_2 = E + \varepsilon$, namely, the energy levels of the side dots are aligned and that of the center dot is detuned (see Fig. S1(a)). Next we introduce finite tunnel coupling between neighboring dots described by the Hermitian matrix,

$$T = \begin{pmatrix} & t_{12} & \\ t_{21} & & t_{23} \\ & t_{32} & \end{pmatrix}, \quad t_{12} = t_{21}^*, \quad t_{23} = t_{32}^*. \quad (A.2)$$

Here we fix the tunnel couplings as $t_{12} = t_{23} = t$ for simplicity. Thus the total Hamiltonian can be written in the matrix form,

$$H = H_0 + T = \begin{pmatrix} E & t & \\ t & E+\varepsilon & t \\ & t & E \end{pmatrix}, \quad (A.3)$$

with modified eigenvalues,

$$E_0 = E, \quad E_\pm = E + \frac{\varepsilon \pm \sqrt{\varepsilon^2 + 8t^2}}{2}, \quad (A.4)$$

and corresponding eigenstates, $|\phi_0\rangle = (|\phi_1\rangle - |\phi_3\rangle)/\sqrt{2}$, $|\phi_+\rangle$ and $|\phi_-\rangle$, respectively. In the presence of finite inter-dot tunnel coupling, the original eigenstates are hybridized and molecular orbitals $|\phi_+\rangle$ and $|\phi_-\rangle$ are formed although $|\phi_0\rangle$ has no contribution from $|\phi_2\rangle$ (see Fig. S1(b)).

We plot these eigenvalues as a function of detuning $\varepsilon$ in Fig. S1(c). At $\varepsilon = 0$, all the eigenstates are fully delocalized. As we detune the energy level of the center dot, however, either $E_+$ or $E_-$ approaches $E_0$ and the eigenstates gradually become localized. For large positive or negative detuning ($|\varepsilon| \gg t$), $E_\pm \approx \pm 2t^2/\varepsilon$ is satisfied and $|\phi_0\rangle$ and $|\phi_\pm\rangle$ are almost degenerate. In these conditions tunneling between the side dots via the center dot is no longer significant. Using (A.4) the indirect tunnel coupling energy $E_t$ can be defined as the energy difference between $|\phi_0\rangle$ and an energetically closer eigenstate, which can be represented by

$$E_t = \frac{\sqrt{\varepsilon^2 + 8t^2} - |\varepsilon|}{2}. \qquad (A.5)$$

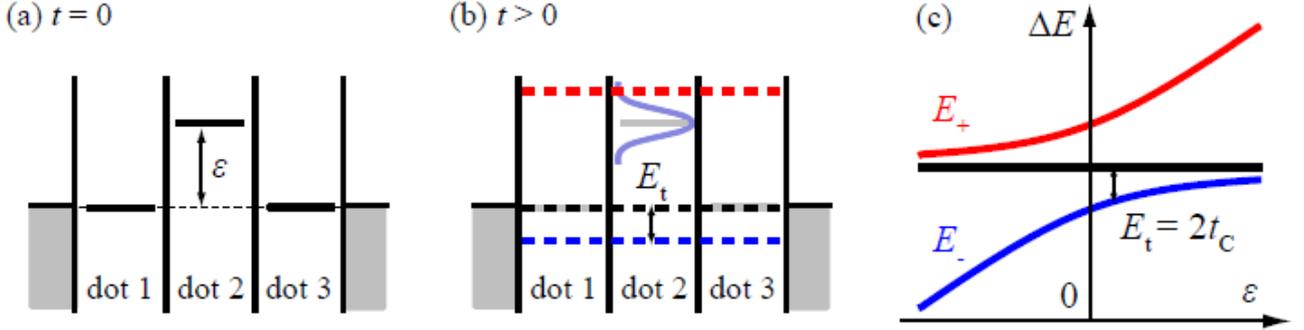

B. Supplementary figures
S1. Energy-level diagrams of the TQD for (a) t = 0 and (b) t > 0, respectively. We assume that the energy levels of the left and the right dots are aligned while that of the center dot is detuned by $\varepsilon$. Finite inter-dot tunnel couplings yield the molecular orbitals. (c) Eigenvalue of each orbital as a function of $\varepsilon$. We define the inter-dot tunnel coupling energy $E_t$ as shown in the energy diagram.